\documentclass[twocolumn,showpacs]{revtex4}

\usepackage{graphicx}
\usepackage{amssymb,amsfonts,amsmath,psfrag,latexsym}

\def\be{\begin{equation}}
\def\ee{\end{equation}}
\def\bea{\begin{eqnarray}}
\def\eea{\end{eqnarray}}
\def\a{\alpha}
\def\b{\beta}

\def\be{\begin{equation}}
\def\ee{\end{equation}}
\def\bea{\begin{eqnarray}}
\def\eea{\end{eqnarray}}
\def\a{\alpha}

\begin{document}

\title{Space-time approach to  microstructure selection  in solid-solid transitions}

\author{Surajit Sengupta$^{1,2}$, Jayee Bhattacharya$^{2}$,Madan Rao$^{3,4}$}
\affiliation{$^1$ Centre for Advanced Materials, Indian Association for the Cultivation of Science, Jadavpur, Kolkata 700032, India\\$^{2}$Advanced Materials Research Unit, S. N. Bose National Centre for Basic Sciences, Salt Lake, Kolkata 700091, India\\$^{3}$ Raman Research Institute, C.V. Raman Avenue, Bangalore 560080, India \\$^{4}$ National Centre for Biological Sciences (TIFR), Bellary Road, Bangalore 560065, India}


\begin{abstract}
Nucleation of a solid in solid is initiated by the appearance of distinct dynamical heterogeneities, consisting of
`active' particles whose trajectories show an abrupt transition from ballistic to diffusive,
coincident with the discontinuous transition in microstructure from a {\it twinned martensite} to {\it ferrite}. The active particles exhibit intermittent jamming and flow. The nature of active particle trajectories decides the 
 fate of  the  transforming solid -- on suppressing single particle diffusion, the transformation proceeds via rare string-like correlated excitations, giving rise to twinned martensitic nuclei. 
 We characterize this transition using a thermodynamics in the space of trajectories in terms of a dynamical action for the active particles confined by the inactive particles. 
\end{abstract}


\maketitle

A quench across a solid-solid transformation generally results in a product solid with specific microstructure -- a long-lived, mesoscale ordering of atoms \cite{cahn,rob} driven far from equilibrium. Crystallographic mismatch, usually present at the growing product-parent interface needs to accommodated dynamically to preserve continuity. This may happen in a variety of ways, giving rise to myriads of possible microstructures depending on the quench protocol (history). Commonly characterized microstructures like {\it ferrite} and {\it twinned martensite}  \cite{cahn,rob,kaushik} differ greatly. While ferrite is associated with a disorderly (or ``civilian'') movement of atoms, martensite is characterized by a coordinated (or ``military'') motion\cite{cahn,rob}, often resulting in alternating variants of the product sharing a common crystallographic mirror plane (twins) \cite{kaushik}. Relating multi-scale physics, from atomic trajectories to mesoscale ordering, makes the study of microstructure selection particularly challenging. Here, we show that accommodation of interfacial mismatch occurs by the appearance of dynamical heterogeneities in the transforming solid.  A thermodynamics of space-time trajectories of active particles\cite{Chandler1,Chandler2}, represented by a dynamical action\cite{edwards,cohen} is used to  describe a sharp transition in microstructure resembling ferrite to martensite as single particle diffusion is suppressed.  The distribution of a space-time order parameter characterizing the nature of trajectories of active particles shows an abrupt change coinciding with this microstructural transition. Active particles exhibit intermittent jamming and flow showing that the underlying physics shares common features with the physics of plasticity\cite{Langer1,Maloney}, glass\cite{Chandler1,Langer2} and granular systems\cite{Jamming1,Jamming2,Jamming3}.

\begin{figure*}[ht]
\begin{center}
\includegraphics[width=5.0 in]{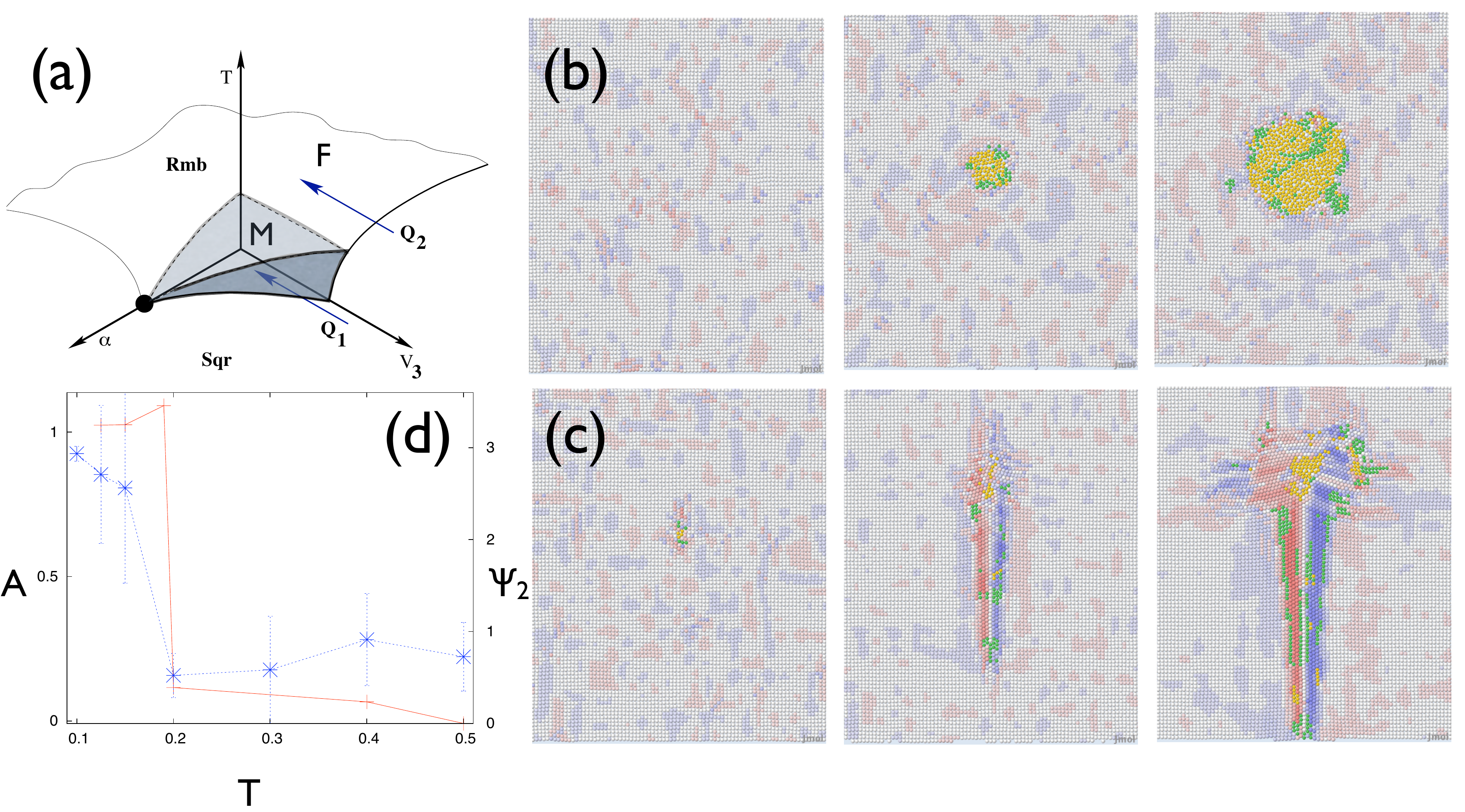}
\caption{(a) Phase diagram in $T-v_3-\alpha$ at $\langle \rho \rangle = 1.05$, indicating the {\it equilibrium} $P4m \to P2$ (Sqr$\to $Rmb) first order phase transition (curved surface bounded by solid lines) as well as the 
{\it dynamical transition} between ferrite ($F$) and martensite ($M$) (shaded surface). 
Particles interact via an effective (purely repulsive) short range potential - a sum of an anisotropic 2-body 
 potential (anisotropy parametrized by $\alpha$) and a 3- body 
 potential, whose strength is parametrized by $v_3$\cite{jphys1,epaps}. The large black dot on the $\alpha$ axis denotes a tricritical point where the jump in the order parameter $e_3$ vanishes.  
  (b) Snapshots of the growing $F$ nucleus at temperature $T=0.7$ following the quench ($Q_2$, see (a)) to the $F$-phase at times $t = .1, 1,$ and $5$ (all units defined in\cite{epaps}). (c) Snapshots of the growing twinned nucleus following the quench ($Q_1$, see (a)) to the $M$-phase at $T=0.1$ at comparable values of $t$ . The colors denote $e_3$ and the nonaffine parameter $\phi$ \cite{jphys1,epaps}, Red $\to e_3 > 0$, Blue $\to e_3 < 0$, Gold $\to \phi > 0$ and Green $\to \phi < 0$. 
(d) {\it Shape anisotropy} $A $ (blue $\ast$, axis on the left) and the {\it degree of twinning} $\Psi_2$ (red $+$, axis on the right) of the product nucleus, (for definitions see \cite{epaps}) vs.  $T$, shows an abrupt jump across the dynamical transition. 
}
\label{dpd}
\end{center}
\end{figure*}

Our model two dimensional (2d) solid\cite{prl2}, analyzed using a molecular dynamics (MD) simulation\cite{SMIT}, consists of particles confined within  a box and interacting via an effective
potential which is a sum of an anisotropic  2-body and 3- body potentials\cite{RScondmat}, constructed so as to 
produce a $P4m \to P2$ (square$\to $rhombus) transition as the temperature or strength of the 
potential is varied at fixed density, Fig.\,\ref{dpd}\,a. The order parameters for this {\it equilibrium} transition are the shear and deviatoric strains $(e_3,e_2)$.  Our earlier MD work on microstructure selection\cite{jphys1}, addressed the issue of dynamical selection from a more mesoscopic point of view. Though we focus on a model solid-solid transformation, our conclusions should hold very generally.

A ``quench'' from the square parent phase across the structural transition, nucleates rhombic regions\,; the shapes, microstructure, and nucleation dynamics  of  product nuclei depend on the
`depth-of-quench' (Fig.\ref{dpd}a). A quench
into region $F$ (ferrite), results in a critical nucleus which is isotropic,
untwinned and  with a rhombic  microstructure separated by grain boundaries (Fig.\ref{dpd}b). In contrast, a 
quench into region $M$ (martensite), nucleates 
a `needle-like' region -- the critical
nucleus is highly anistropic, with the long axis lying along one of the axes of the parent square
lattice (Fig.\ref{dpd}c). The microstructure is twinned --- a plot of
the shear strain $e_3$ reveals that it changes sign across a twin interface between two degenerate product variants  which lies along one of the square axes.
 The regions $F$ and $M$ are separated by a {\it dynamical phase boundary}, as seen from the 
sharp  jumps of (i) the shape anisotropy \cite{jphys1} and (ii) the degree of twinning of the critical nucleus, at the transition (Fig.\ref{dpd}d). In both phases, the nucleation of the product is accompanied by the formation of nonaffine zones (NAZ) characterized by a large value of the nonaffine parameter
$\phi$, derived from elastic strains by coarse-graining the particle displacements \cite{epaps} --
the $M$ and $F$ nuclei exhibit a distinct patterning and dynamics of the NAZs \cite{jphys1}. 
We emphasize that this dynamical phase boundary is sharp only when the quench rate is infinitely fast.

 We now look for signatures of this dynamical transition in the microscopic trajectories of particles. We find that the nucleation and growth of the product is initiated by the movement of a fraction of particles, which we term {\it active}. The clusters formed by these active particles define 
{\it dynamical heterogeneities} in the transforming solid. We find that the dynamical heterogeneities so defined,
overlap completely with the more coarse grained NAZs. 
\begin{figure}[ht]
\begin{center}
\includegraphics[width= 3.5in]{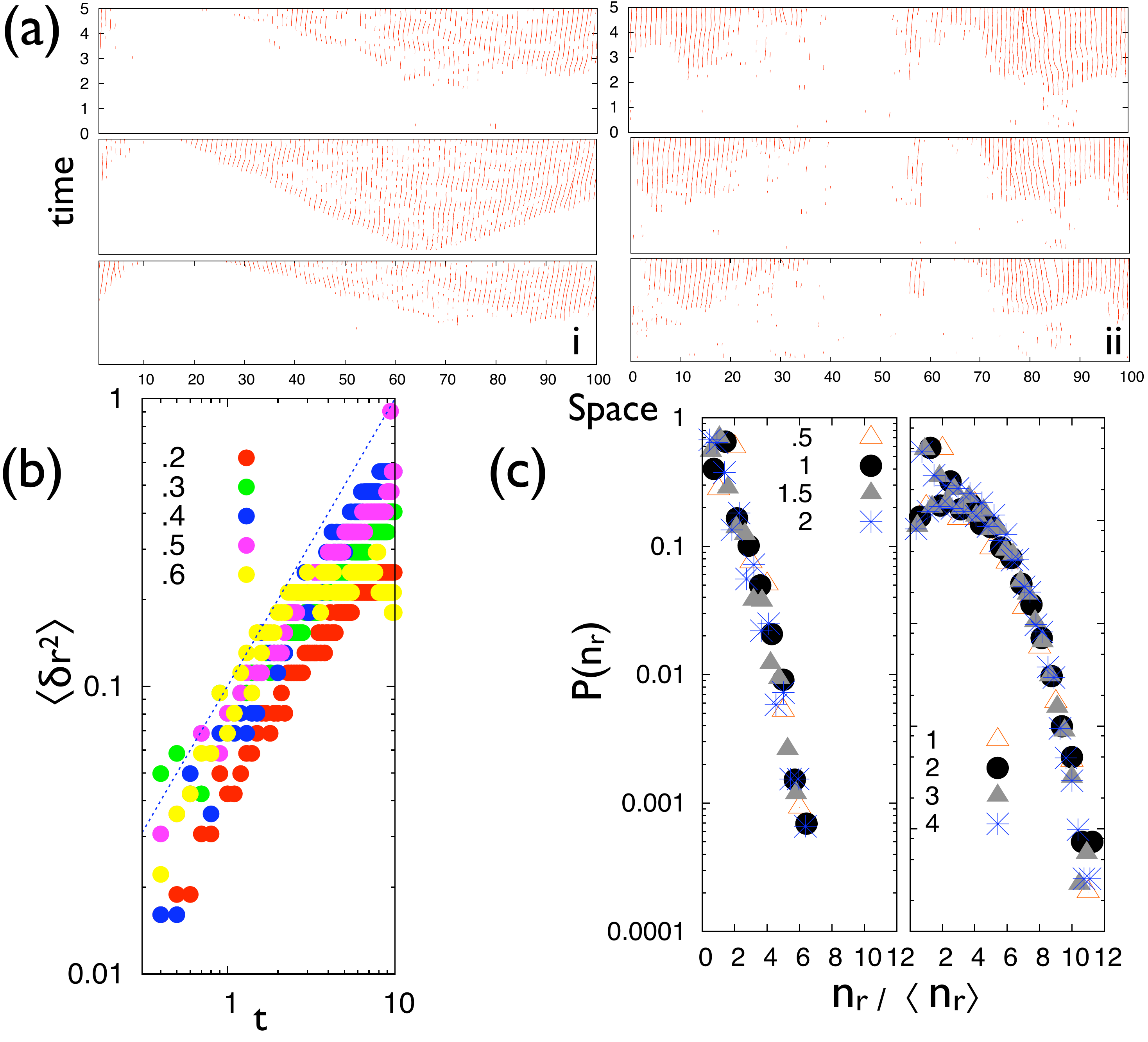}
\caption{  (a) Space-time {\em kymographs} denoting particle trajectories, i.e., plots of $y(t)$ for particles at fixed values of $x$, showing clusters of active particles surrounded by inactive particles (empty patches). (i) Kymographs in the $M$-phase, $T=0.1$, at $x=10, 15, 20$ (top to bottom panels) and (ii) Kymographs in the $F$-phase, $T=0.6$, at $x=15, 26, 57$ (top to bottom). 
 Active particles in $M$ are concentrated near the edge of the single growing nucleus, while
in $F$, they are distributed throughout the sample. The movement of active particles is ballistic in $M$ and random in $F$\cite{epaps}. The alternate cycles of (in)activity and the collective active (unjamming) $\to$ inactive (jamming) are apparent in the kymographs. (b) Tagged particle diffusion coefficient, obtained from the slope of $\langle \delta r^2 \rangle = \int_0^{t} \sum_{i} ({\bf r}_i(t') - {\bf r}_i(t' - \Delta t'))^2 dt' $ vs. $t$ with $\Delta t'= 0.1$ (sum is over all active particles), is independent  of  temperature, over the range $T = .2-.6$ in the $F$-phase. (c) Statistics of (in)active interconversions in terms of the probability distribution $P(n_r, t)$ for  the number of reversals $n_r$ (experienced by all particles from active to inactive and vise-versa) upto time $t$, plotted against $n_r$, for different times
($t = 0.5 - 2.0$ for the $M$-phase at $T=0.1$ (left), and  $t =  1.-4.$ for the $F$-phase at
$T=0.6$ (right)). The data collapse shows that $n_r/\langle n_r \rangle$ is the scaling variable, where 
$\langle n_r\rangle$ is the mean number of reversals upto time $t$. In both phases,  $P(n_r, t)$ decays exponentially, with a possible stretching in the $F$-phase.}
\label{kymo}
\end{center}
\end{figure}

A closer look at the space-time trajectories within and in the vicinity of the dynamical heterogeneities in the form of 
 {\it kymographs} (Fig.\ref{kymo}a), reveals  fine differences between the $M$ and $F$ -phases.
In the M-phase, we see that nucleation is accompanied by the birth of a small fraction of active particles, surrounded by regions of inactive particles \cite{Chandler2}. At these early times they move ballistically and in
a coordinated manner; the velocity correlation is significant both along a single trajectory and across neighboring trajectories at equal time.  The number of active particles grows with time, with the older active particles seeding newer ones. Note that the highly coordinated military movement,
characteristic of a martensite, is apparent in these kymographs (Fig.\ref{kymo}a and \cite{epaps}).
In the $F$-phase, the fraction of active particles is larger and their trajectories intersect  many times\cite{epaps}; the single particle trajectories are diffusive over time scales larger than
the velocity correlation time. Indeed we find that the tagged-particle diffusion coefficient
 is  large and athermal (Fig.\ref{kymo}b) and is driven by local variations in 
 free-volume as a result of accomodation \cite{jphys1}. 
\begin{figure}[h]
\begin{center}
\includegraphics[width=2.5in]{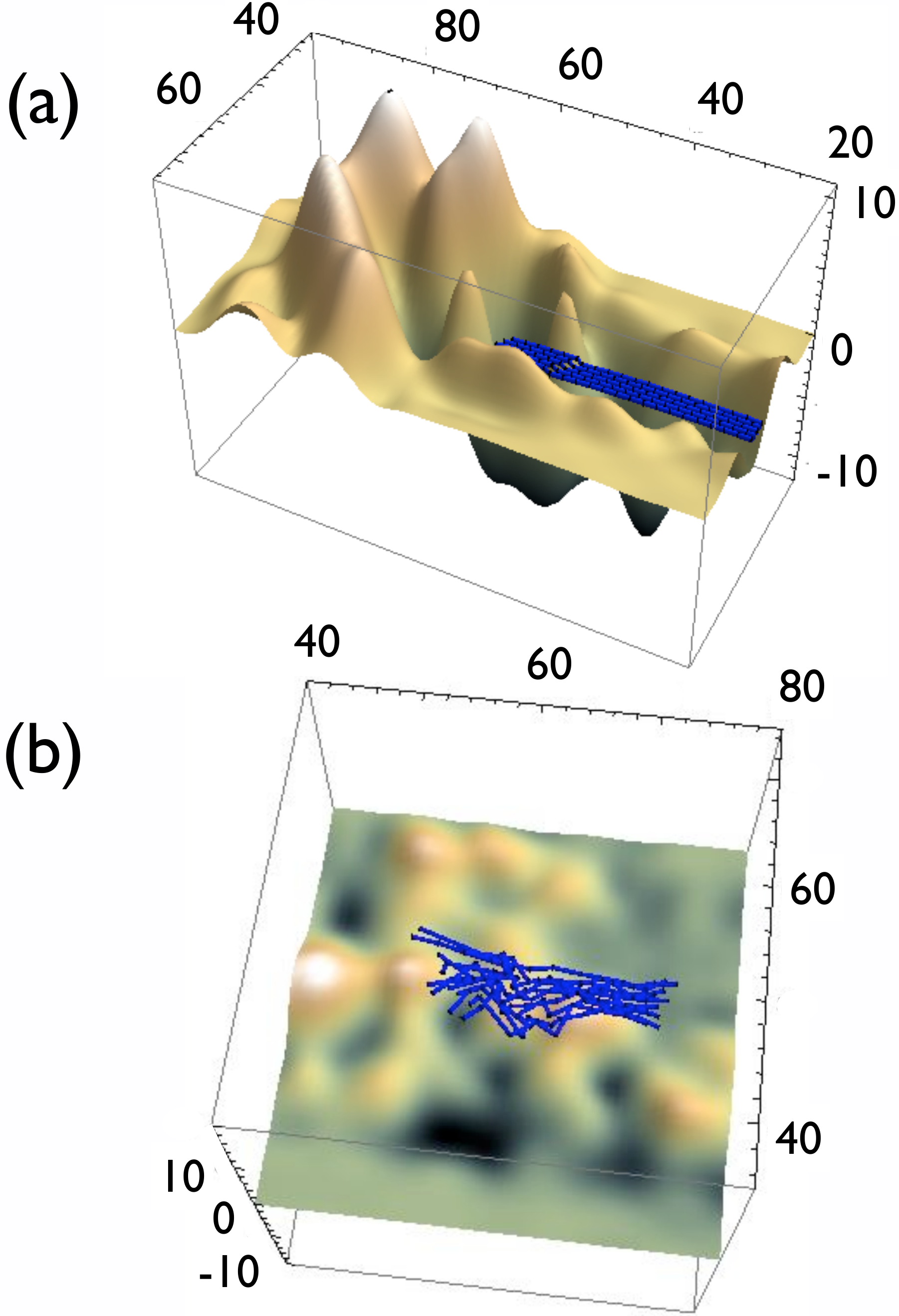}
\caption{Topography of local energy landscape due to the inactive particles, showing deep ridges and shallow delta in the $M$ (a) and $F$ (b) phases, respectively. The energy landscape is obtained by using the fitted local strain $e_3(x,y)$ in the usual (local) nonlinear elastic free energy density $a e_3^2 - b e_3^4 + c e_3^6$, where coefficients $a, b, c$ are proportional to the elastic moduli obtained explicitly from total energy calculations (see \cite{RScondmat}, for details). In (a) the deep ridge directs rows of active particles (blue lines) along a specific channel.  (b) In the $F$-phase, the active particle trajectories span out isotropically. Thus the collective excitations are string-like in the $M$-phase (a) and diffusive in the $F$-phase
(b). }
\label{topo}
\end{center}
\end{figure}

The active particle  trajectories show alternate arrest and movement, both in the $F$ and $M$ phases (Fig.\ref{kymo}b), thus particles continuously transform from an active to inactive state.  We study the statistics of such activity transitions in the two phases; the distribution of these state reversals 
(Fig.\ref{kymo}c)  is exponential, with a possible stretching in the $F$-phase.
Note that the arrest of all active particles within a dynamical heterogeneity happens roughly simultaneously (Fig.\ref{kymo}c). This is the space time realization of the observed jamming and flow\cite{Maloney} in the stress-flow curves within the NAZs \cite{jphys1}.

The active particles in  the $M$-phase are dynamically hindered and do not explore local configuration space. This is dramatically apparent in the topography of the local energy landscape set by the inactive particles, which we compute by using the fitted local order parameter, $e_3(x,y)$ in a nonlinear elastic free energy density (Fig.\ref{topo}) -- this shows deep ridges in the local free-energy landscape, which herd active particles along a narrow channel, creating string-like excitations\cite{Langer2}. The width of the string is of the order of the lateral scale of the dynamical heterogeneities.
 In the F-phase, the landscape exhibits many criss-crossing shallow ridges which directs particles here and there, ending up in a large scale topography akin to a delta. The delta spans  out isotropically, leading to diffusive collective excitations. Thus the difference in the space-time trajectories arises from the kinetic constraints on active particles created by the energy landscape due to inactive particles.
\begin{figure}[t]
\begin{center}
\includegraphics[width=3.0 in]{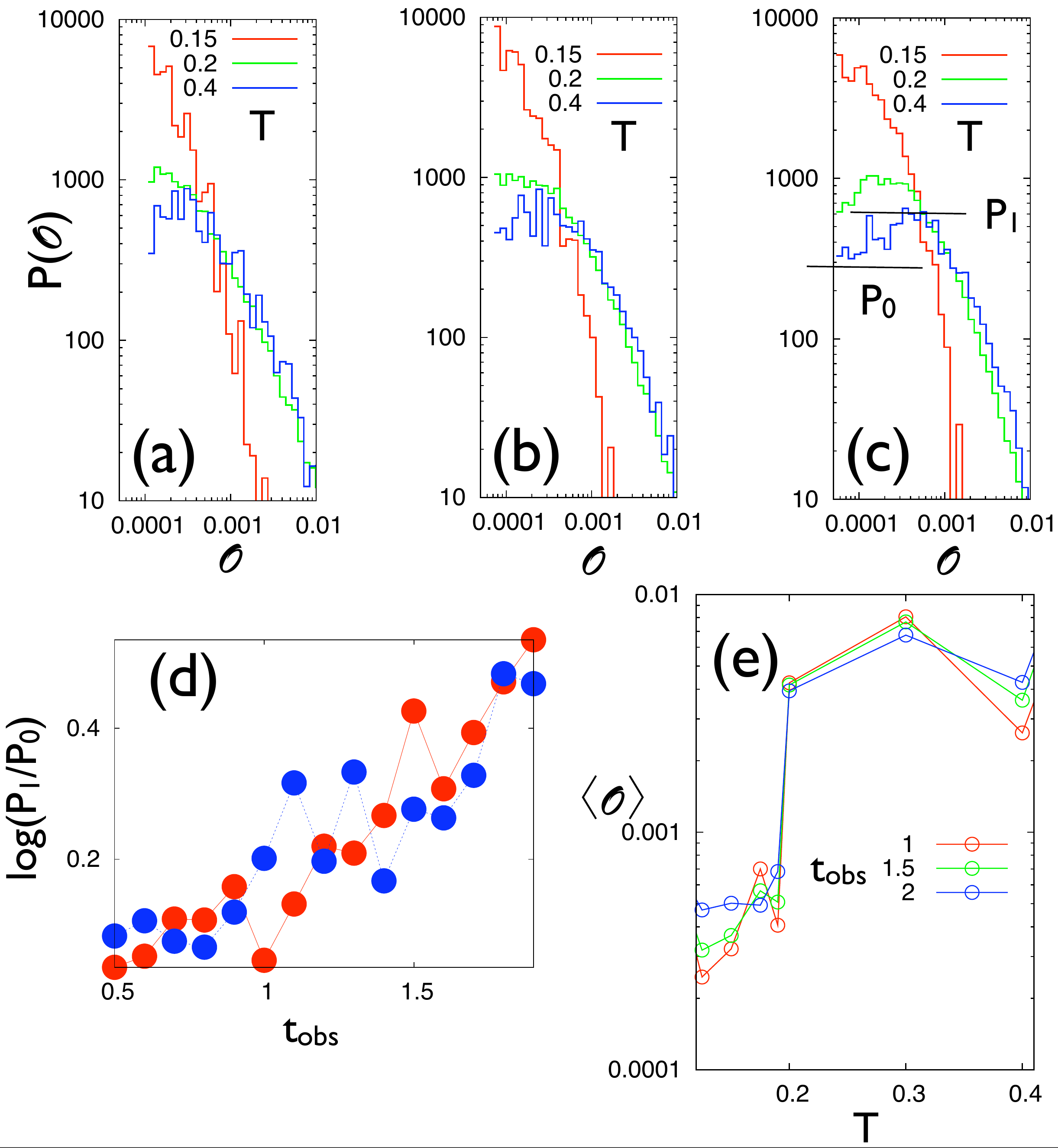}
\caption {(a) -(c) Probability distribution of the off-diagonal order parameter $P({\mathcal O})$ for $T=0.15, 0.2$ and $0.4$ for $t_{obs} = 1$ (a), $1.5$ (b) and $2$ (c). Note the development of a peak at ${\mathcal O}_{max} \neq 0$ for $T > 0.2$, as $t_{obs}$ increases. (d) Finite-size scaling in space-time, measured by $\log(P_1/P_0)$ versus $t_{obs}$, where $P_1$ is the probability at the maximum value ${\mathcal O}_{max}$ and $P_0 = P(0)$ shown for two temperatures $T = 0.3$ (blue circles) and $0.4$ (red circles). The transition sharpens with increasing $t_{obs}$. (e) Mean $ \langle {\mathcal O} \rangle$ as a function of $T$ for $t_{obs} =  1, 1.5$ and $2$ shows a sharp transition at 
 $T\approx0.2$ (compare Fig.\ref{dpd}d).}
\label{scaling}
\end{center}
\end{figure}

To quantify the abrupt changes in the nature of trajectories of (a
few) active particles moving in the confining potential created by the
(majority) inactive particles, we take a space-time approach\cite{Chandler1,Chandler2} and define a
thermodynamics of phase transitions in trajectory space.
The Lagrangian displacements $u_{i\a}(t) =
r_{i\a}(t)-r_{i\a}(t-\delta t)$, where
$\a = 1, 2$ (along x and y directions for the square-to-rhombus
transition) and $\delta t$ is a small time offset.
The active particles move `freely' in the confining potential set by
the inactive particles, thus the probability
distribution of trajectories of active particles is given by
 $P\left[ {\cal C}\right] \propto \exp\left( - S\left[  {\cal C}
\right]\right)$\cite{edwards,cohen}, where the `stochastic  action' is given by,
\be
S\left[ {\cal C}\right]  =  \int_0^{t_{obs}}  \frac{1}{2D}
\sum_{i,\a} ( \tau_m {\ddot u}_{i\a} + {\dot u}_{i\a})^2\,  dt\, ,
\ee
upto an observation time, $t_{obs}$. Here $D$ is a diffusion coefficient and the summation is over active particles
 alone subject to the constraint that the inactive particles make
large portions of configuration space
 inaccessible \footnote{There is a further constraint : in the
absence of external forces, the centre of mass displacement
of the system should be zero.}. This constraint can be expressed as an
effective confining potential experienced by the active particles,
which augments the `free particle' action by a term of the form,
 $- \int_0^{t_{obs}} \sum_{i\a\b} \left[V_{\a\b}u_{i\a}(t) u_{i\b}(t)
+ \ldots \right]\, dt$.

The two phases $F$ and $M$ are characterized by the  space-time
intensive off-diagonal order parameter, constructed from the bilinear
$\Delta^i_{\a\b}(t) = u_{i\a}(t)  u_{i\b}(t)$ with $\a \neq \b$,
\bea
{\mathcal O}/N & \equiv  & \frac{1}{t_{obs} N} \int_{0}^{t_{obs}}\, dt \, \sum_{i}
\vert \Delta^i_{\a\b}(t)\vert^2 \, .
\label{orderparam}
\eea
The typical value of ${\mathcal O}/N$  is zero in the $M$-phase and
undergoes a sharp jump
of $O(1)$ in the $F$-phase (Fig.\ref{scaling}a-c). The order parameter probability
distribution $P({\mathcal O})$, is exponential in the $M$-phase
with a peak at ${\mathcal O} = 0$, and changes  over to a distribution with a peak
at ${\mathcal O}_{max} \neq 0$ in the $F$-phase. 
To declare this a bona-fide space-time phase transition, the distribution $P({\mathcal O})$
should exhibit  ``finite-size'' scaling, with respect to changes
in the space-time ``volume'' $N t_{obs}$. This is most apparent in the scaling behaviour of 
$\log(P_1/P_0)$ (Fig.\ref{scaling}d), where  $P_0 = P(0)$ and $P_1 = P({\mathcal O}_{max})$. 
In an equilibrium Ising phase transition, the analogous quantity measures the free-energy of domain-walls and increases as $L^{d-1}$, where $L$ is the linear system size and $d$ the dimensionality
of space\cite{Binder}. Figure \ref{scaling}d shows that $\log(P_1/P_0)$  increases as $t_{obs}$ 
increases,  and thus  scales with the ``volume'' of space-time, indicating that the 
$M-F$ transition becomes sharper in the space-time thermodynamic limit. The mean $\langle {\mathcal O} \rangle$ shows a sharp jump (Fig.\ref{scaling}e) at the same temperature where the structural order-parameters at the mesoscopic scale, viz., $A$ and $\Psi_2$, show sharp changes (Fig.\ref{dpd}d).

Our work provides the physical link between
microscopic particle trajectories and mesoscopic microstructure of the product solid, mentioned in the
introduction. Our study on the dynamics of solid nucleation also reveals hitherto unsuspected connections with the physics of glass\cite{Chandler1,Chandler2,Langer2}, granular media and jamming\cite{Jamming1,Jamming2,Jamming3} and plasticity\cite{Langer1,Maloney}.
We hope to pursue the many ramifications that these
links promise, e.g.,  the spatiotemporal statistics of dynamical heterogeneities leading to intermittent jamming and flow; developing a nucleation theory of solid state transformations arising from string-like excitations; and formulating an explicit Landau theory to describe the thermodynamics of phase transitions in trajectory space.

\begin{acknowledgments}
SS and MR acknowledge support from DST (India) and HFSP, respectively.
\end{acknowledgments}

\end{document}